\documentclass[prl,superscriptaddress,twocolumn]{revtex4-1}
\usepackage{times,amsmath}
\usepackage{epsfig}
\usepackage{graphicx}
\usepackage{dcolumn}
\usepackage{bm}
\usepackage{tabularx}
\usepackage{hyperref}
\hypersetup{citecolor=black, filecolor= black, linkcolor= black, urlcolor= black}
\begin{document}

\preprint{version 0}

\title{Evolutionary Method for Predicting Surface Reconstructions with Variable Stoichiometry}

\author{Qiang Zhu}
\email{qiang.zhu@stonybrook.edu}
\affiliation{Department of Geosciences, Stony Brook University, Stony Brook, New York 11794, USA}
\author{Li Li}
\affiliation{Department of Physics and Astronomy, Stony Brook University, Stony Brook, New York 11794, USA}
\author{Artem R. Oganov}
\affiliation{Department of Geosciences, Stony Brook University, Stony Brook, New York 11794, USA}
\affiliation{Department of Physics and Astronomy, Stony Brook University, Stony Brook, New York 11794, USA}
\affiliation{Geology Department, Moscow State University, Moscow 119992, Russia}
\author{Philip B. Allen}
\affiliation{Department of Physics and Astronomy, Stony Brook University, Stony Brook, New York 11794, USA}
\date{\today}

\begin{abstract}
We present a specially designed evolutionary algorithm for the prediction of surface reconstructions. This new technique allows one to automatically explore all the low-energy configurations with variable surface atoms and variable surface unit cells through the whole chemical potential range. The power of evolutionary search is demonstrated by the efficient identification of diamond 2$\times$1 (100) and 2$\times$1 (111) surfaces with a fixed number of surface atom and a fixed cell size. With further variation of surface unit cells, we study the reconstructions of the polar surface MgO (111). Experiment has detected an oxygen trimer (ozone) motif (Plass et al, 1998). We predict a new version of this motif which can be thermodynamically stable at extreme oxygen rich condition. Finally, we perform a variable stoichiometry search for a complex ternary system: semi-polar GaN $(10\overline{1}1)$ with and without adsorbed oxygen. The search yields a non-intuitive reconstruction based on N$_3$-trimers. These examples demonstrate that an automated scheme to explore the energy landscape of surfaces will improve our understanding of surface reconstructions. The method presented in this report can be generally applied to binary and multi-component systems.
\end{abstract}
\maketitle

\section{Introduction}
At crystalline surfaces, especially in semiconductors, atoms rearrange to form reconstructed configurations, which are different from those in the bulk crystals. Reconstructions play a key role in surface properties, such as epitaxial crystal growth, catalysis, etc. Determining the reconstruction of semiconductor surfaces is a long-standing problem. Despite progress in experimental techniques, it is still hard to obtain the atomic-scale details. Diffraction and microscopy can yield the periodicity and symmetry \cite{Xue-PRL-1999, Reinert-Inor-2013}. However, the sampling space is astronomically large, and solutions may defy intuition, even for monoatomic systems: consider Si(111) 7$\times$7 \cite{Schlier-JCP-1959, Takayanagi-Surf-1985, Brommer-PRL-1992}. For binary and more complex cases, a variable stoichiometry at the surface adds a huge extra dimension. 
The problematic complexity of surface structures is further compounded by a range of synthesis techniques.  Varying precursor compounds of possibly mixed gas, liquid, or solid phases, varying pressures and temperatures during synthesis, varying annealing schedules, etc., may allow a significant range of metastable surface phases to survive.  Thus experiments may correctly offer conflicting suggestions about structure.  Therefore theory has a role not just to attempt to predict true thermodynamically stable structures, but also to catalog potential metastable phases, needed for guidance in interpretation of experiment.  

From the computational perspective, the main difficulty lies in the fact that the number of candidate structures is extremely large. It is not possible to enumerate all the reasonable reconstruction models just from chemical intuition. There have been a few efforts towards automatic prediction of the most stable configuration for either a given stoichiometry \cite{Chuang-Surface-2004, Briggs-PRB-2007, Henkelman-JPCC-2009}, or a variable composition at constant chemical potential \cite{Sierka-JCP-2007}. With these structural models available, the thermodynamical phase diagram can be constructed \cite{Reuter-PRL-2003}. However, unbiased prediction of reconstructions with variable stoichiometry in the whole chemical potential range has not been attempted to our knowledge.  Here we demonstrate that the unbiased nature of an evolutionary search is able to find unexpected structures that compete thermodynamically with ones already suggested by prior searches. In this paper, we propose a new method which allows automatic exploration of all low-energy configurations, with variable surface atoms and variable reconstruction cells through the whole chemical potential range. We apply this method to a series of systems: diamond (100) and (111), MgO (111) and GaN $(10\overline{1}1)$ surfaces with and without added oxygen.

\section{Theoretical Background}
\subsection{Surface Energy and Chemical Potential }
Relative stability among candidate surfaces is determined by the formation energy $E_\text{formation}$
\begin{equation}
E_\text{formation} = E_\text{tot} - E_\text{ref}-\sum_i n_i \mu_i,
\label{eq:E1}
\end{equation}
where $E_\text{tot}$ and $E_\text{ref}$ are the total energy of the surface under consideration and of the reference cleaved surface; $n_i$ and $\mu_i$ are the number and chemical potential for each species, respectively. The chemical potential is the energy to add or remove one atom from the system, assuming there is a reservoir for each species to equilibrate with. The chemical potentials must satisfy constraints. For a simple binary AB compound, if $\mu_\text{A}$ is extremely high, the elmental phase A would condense on the substrate AB. Therefore, the chemical potential of A (or B) is upper bounded by the chemical potential of elemental A (or B),
\begin{equation}
\begin{array}{rcl}
\mu_\text{A} \leq \mu^{0}_\text{A},\\
\mu_\text{B} \leq \mu^{0}_\text{B}.
\end{array}
\end{equation}
The A and B atoms are in equilibrium with the substrate,
\begin{equation}
\mu_\text{A} + \mu_\text{B} = G_\text{AB}.
\label{eq:AB}
\end{equation}
At $T$ = 0 K, the Gibbs free energy $G_\text{AB}$ can be simplified to the internal energies $E_\text{AB}$.

These conditions lead to the range of the A component chemical potential to be
\begin{equation}
E_\text{AB} - \mu^{0}_\text{B} \leq \mu_\text{A} \leq \mu^{0}_\text{A}.
\label{eq:Ga}
\end{equation}
Therefore, Eq.~(\ref{eq:E1}) can be rewritten as only $\mu_{\text{A}}$ dependent
\begin{equation}
E_\text{formation} = E_\text{tot} - E_\text{ref} - n_{\text B}E_\text{AB} - \mu_\text{A} (n_\text{A} - n_\text{B}).
\label{eq:E2}
\end{equation}
The surface phase diagram shows which reconstruction is favored under a specific environment, i.e., temperature and chemical potential or partial pressure of each component.  Each of these environmental conditions represents an independent dimension in the phase diagram.

\subsection{Evolutionary algorithm}
We employ an evolutionary algorithm (EA) to explore the surface structures. The algorithm is implemented in the {\small USPEX} package, which has been successfully applied to various bulk materials \cite{Oganov-JCP-2006,Oganov-ACR-2011,Zhu-Acta-2012,Lyakhov-CPC-2013}. We use a representation in which the system has three parts: vacuum, surface and substrate. Vacuum and substrate regions are pre-specified, while the surface region is optimized by the EA. The number of surface atoms varies up to a predefined maximum number for a given thickness of surface. To allow automatic exploration of reconstructions with different surface unit cells, the cell size is also variable.

Our algorithm is constructed in the following way. It first generates random structures with only the surface region being varied. These initial structures are then relaxed, and ranked by fitness, based on the energy. Structures with better fitness are more likely to be selected as parents to generate new child structures. We considered three ways to produce offspring. 

\begin{figure}
\epsfig{file=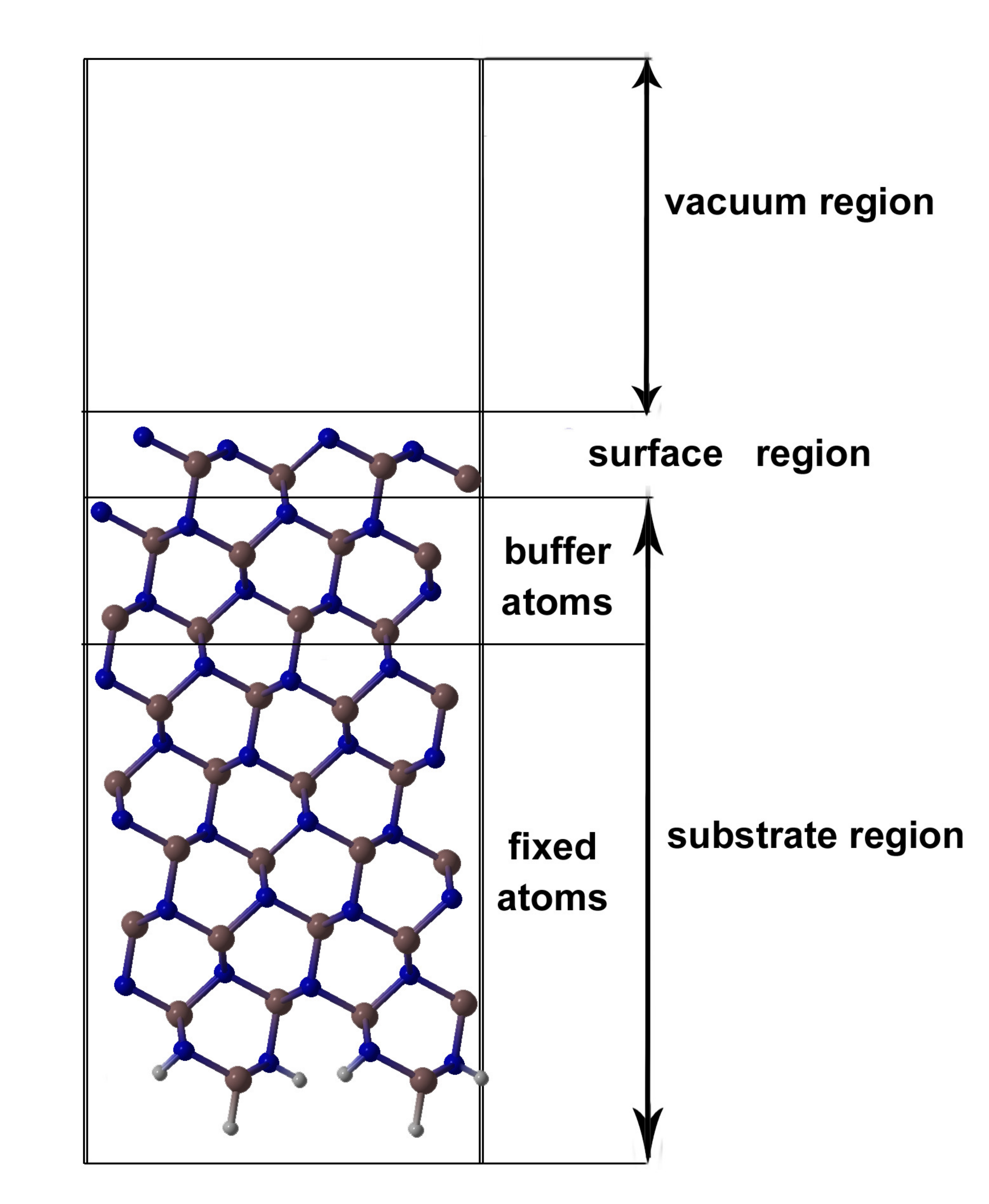, width=0.3\textwidth}
\caption{Surface representation used in the evolutionary algorithm.}
\label{model}
\end{figure}

(i) Heredity: Two structures are chosen from the previous generation.  They are randomly sliced at the same position in the surface unit cell.  Then pieces from both parent structures are combined to generate the offspring. 

(ii) Mutation: One structure is chosen as parent. Similar to the implementation of softmutation in bulk crystals \cite{Lyakhov-CPC-2010}, the surface atoms are displaced according to the softest surface vibrational modes based on the bond-hardness model \cite{Li-PRL-2008, Lyakhov-PRB-2011}. If the structure cannot be softmutated anymore, we switch to the coordinate mutation \cite{Lyakhov-CPC-2010}, which is to randomly displace the surface atoms by an amount drawn from a zero-mean Gaussian distribution and biased by a specific measure of the local order \cite{Oganov-JCP-2009}.

(iii) Transmutation: One structure is chosen as parent. Some atoms are transmuted to another chemical species.  

The offsprings, together with a few best structures from the previous generation, comprise the next generation.  This process is repeated until no lower-energy  structures are produced for sufficiently many generations. 

\subsection{Fitness function}
\begin{figure}
\epsfig{file=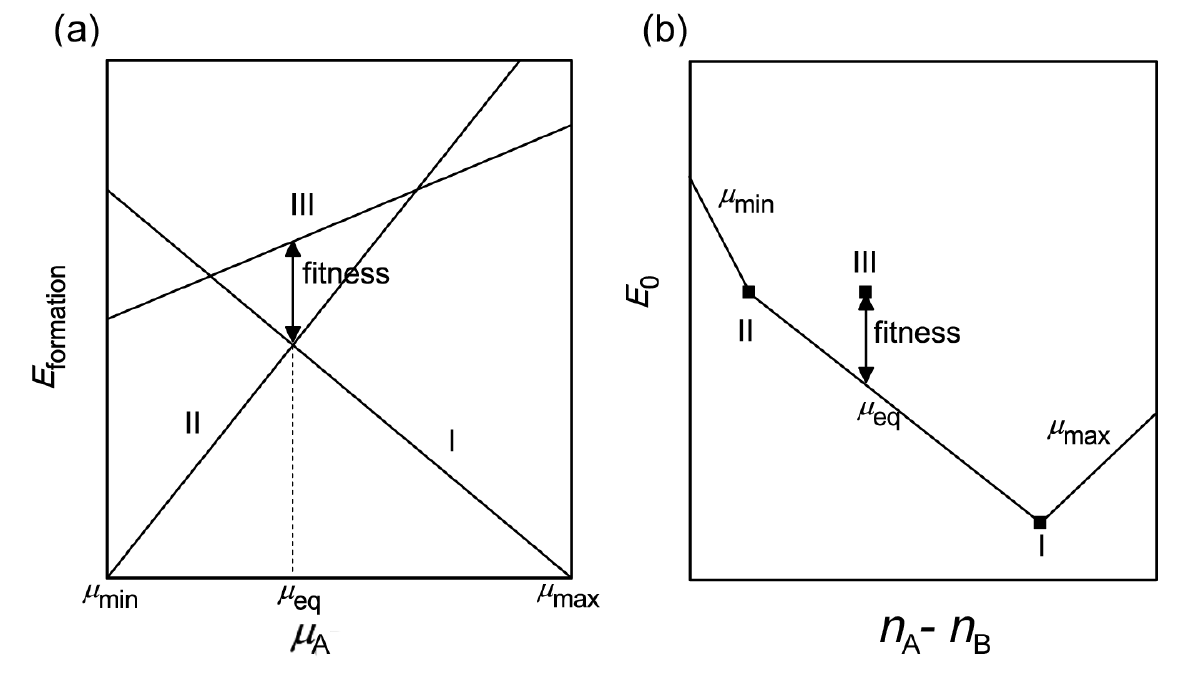, width=0.5\textwidth}
\caption{Illustration of the fitness function used in the evolutionary algorithm. a). Phase diagram as a function of $\mu_\text{A}$. b). Phase diagram as a function of $(n_\text{A}-n_\text{B})$. The vertices of the convex hull are the stable structures appearing in the phase diagram. The slope of each section of the convex hull is either the boundary chemical potential or the equilibrium chemical potential where stable structures can coexist.}
\label{fitness}
\end{figure}

Our fitness function needs to indicate the relative stability of structures with various surface stoichiometry and reconstruction cell sizes. We illustrate this in Fig. \ref{fitness}. First, for two given surface configurations (I and II) which have different numbers of atoms on the same substrate cell, their relative energy difference depends on the chemical potential $\mu_\text{A}$ according to Eq.(~\ref{eq:E2}). As shown in Fig.~\ref{fitness}a, surfaces I and II should coexist at $\mu_\text{A} = \mu_\text{eq}$. Therefore, the stability regime for the two surfaces can be fully established. Surface I is stable when $\mu_\text{min} \leq \mu_\text{A} \leq \mu_\text{eq}$, while surface II is stable when $\mu_\text{eq} \leq \mu_\text{A} \leq \mu_\text{max}$. For any other unstable configuration like surface III, the fitness can be viewed as the minimum energy difference comparing with the stable configuration, at all possible chemical potentials. The minimum condition is reached when $\mu_\text{A} = \mu_\text{eq}$. For practical implementation it is useful to express this algebraically. We define $E_0 = E_\text{tot} - E_\text{ref} - n_{\text B}E_\text{AB}$ as the $\mu_\text{A}$-independent term in Eq.(\ref{eq:E2}). This approach was introduced by Qian {\it et al} \cite{Qian-PRB-1988}, and is widely used \cite{Thomas-PRB-2010}. Versions (a) and (b) of Fig.~\ref{fitness} contain equivalent information. The stable structures appearing on the phase diagram form a convex hull in energy-composition coordinates. The slope of each section of the convex hull is either the boundary chemical potential or the equilibrium chemical potential where stable structures can coexist. We then choose the fitness of a surface structure to be its distance to the convex hull. The EA search then aims to optimize the convex hull, similar to our previously proposed approach of variable composition prediction for bulk crystals \cite{Oganov-ACR-2011, Zhang-arxiv-2012}. When different sizes of surface cell ($m \times n$) occur, the scale factor $N_\text{cell}$ = $m \times n$ should be taken into account and energies have to be normalized.

\subsection{Details of {\it ab initio} calculations}
As a global optimization method, an EA search requires hundreds or thousands of individual structural relaxations. In our calculation, relaxations were done using density functional theory (DFT) within the generalized gradient approximation (GGA) \cite{GGA-1996} using the all-electron projector augmented wave (PAW) \cite{PAW-1994} method as implemented in the {\small VASP} code \cite{VASP-1996}.  We used the kinetic energy cutoff of 550 eV for the plane-wave basis set and Brillouin zone sampling resolution of 2$\pi$ $\times$ 0.08 \AA$^{-1}$, which showed excellent convergence of the energy differences, stress tensors and structural parameters. There are different ways to construct the model. For structures with some kind of mirror symmetry parallel to the surfaces, one can build a symmetric model. The other way is to passivate one surface, for example, by fractional hydrogens \cite{Shiraishi-JPSP-1990}. For a semiconductor this makes the passivated surface semiconducting, with no surface states in the fundamental band gap. Dipole corrections are added in form of a mid-vacuum discontinuity in the electrostatic potential which cancel the interactions between the slab and its periodic images \cite{Neugebauer-PRB-1992, Makov-PRB-1995}. For massive calculations in EA search, we used passivated slab models. Each typical slab contains 5-7 atomic layers and 10-12 \AA ~of vacuum, with the top 2 layers being relaxed. For the post-process, we select all high fitness candidate structures from EA search to construct the surface stability phase diagram. To obtain accurate surface energies, we expand the slab to 8-10 layers, and the vacuum to 12-15 \AA. Both symmetric models and passivation approaches were explored at this stage. Our tests suggest that the surface energy is converged to better than 0.05 eV per surface cell. 

\section{Applications}
In this section, we apply our method to several systems with increasing complexity. First, we study diamond (100)/(111) reconstructions by fixing both variables (number of surface atoms and cell size). For the MgO (111) surface, we focus on the particular stoichiometry (MgO) on the Mg-terminated surface, and explore the possible low-energy reconstructions with a cell size ranging from 1$\times$1 to 2$\times$2. Finally, we investigate the semi-polar GaN $(10\overline{1}1)$ surfaces without and with oxygen absorption, in which the search allows both variable surface atoms and variable surface unit cells. 

\begin{figure*}[!htb]
\epsfig{file=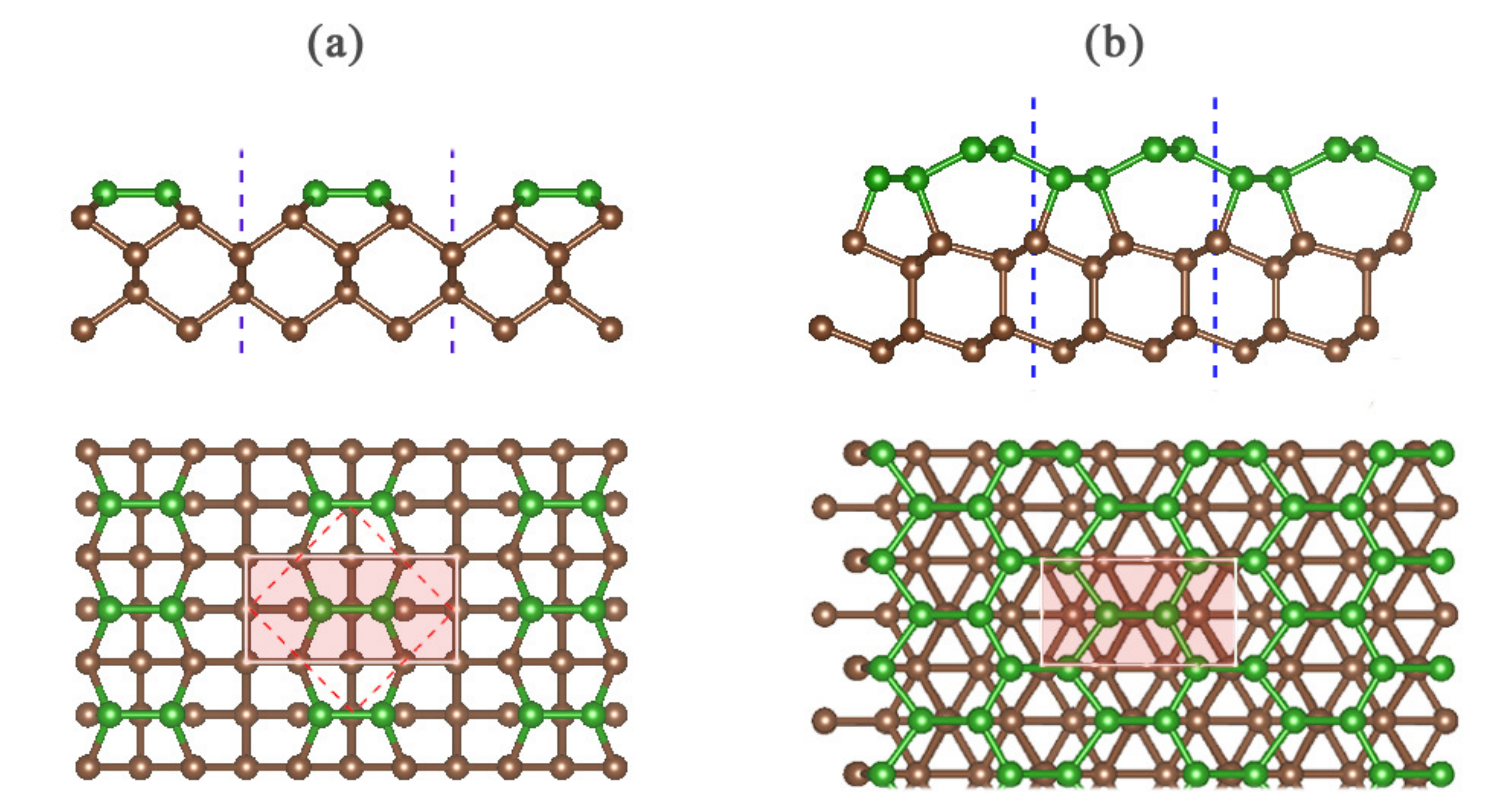, width=0.8\textwidth}
\caption{Schematic image of (a) diamond (100) 2$\times$1 reconstruction (red dashed line: face of conventional cube); (b) diamond (111) 2$\times$1 reconstruction. 2$\times$1 surface unit cells are highlighted by white rectangle.}
\label{diamond}
\end{figure*}

\subsection{Diamond (100) and (111)}

To test our method, we applied the EA search to study the known 2$\times$1 reconstructions of diamond (100) and (111) surfaces, which are the two most important surfaces for polycrystalline diamond from chemical vapor deposition (CVD) \cite{Ristein-surface-2006}. As a benchmark, we fixed the number of surface atoms and cell size in our searches. We tried 2 and 6 carbon atoms on a 2$\times$1 surface cell. During the search, each structure was first relaxed with the Brenner potential \cite{Brenner-PRB-1990} as implemented in the {\small GULP} code \cite{GULP-2003}, and then more accurate energy evaluation was performed at {\it ab initio} level with PBE functional \cite{GGA-1996}. 

Our results are in excellent agreement with those reported in previous literature \cite{Ristein-surface-2006}. 
Fig. \ref{diamond} shows the best structures found in the search. The cleaved diamond (100) surface, containing one unsaturated carbon atom with two dangling bonds per unit cell, is unstable. Stabilization is achieved {\it via} a reconstruction with surface atoms forming one $\pi$-bonded C-C dimer per 2$\times$1 unit cell. This is correctly found by our EA search (Fig. \ref{diamond}a). 
The bulk terminated diamond (111) surface contains 2 unsaturated carbon atoms with 2 dangling bonds per 1$\times$1 unit cell. Pandey first proposed a chain model to explain the Si (111) surface \cite{Pandey-PRL-1981}, and it proved to be correct for diamond as well \cite{Ristein-surface-2006}. Our search also confirmed this model, with surface atoms forming Pandey chains along the $[01\overline{1}]$ direction. From the top view (Fig. \ref{diamond}b.), the Pandey chains further form an extended 2D network, having the same period as the unreconstructed (111) surface. From the side view, the surface atoms together with the second layer, construct an alternating 5+7 ring pattern, which is different from the 6 ring pattern in the bulk. The peculiar 5+7 ring topology also exists in 3D in the recently established structure of a new allotrope of carbon formed by cold compression of graphite \cite{Boulfelfel-SR-2012, Wang-SR-2012}). This demonstrates that either reduced dimensionality or extreme pressure conditions can create a very different chemical environment, leading to very complex reconstructions. Complete searching of these systems' landscapes is needed to find the new chemistry.  

We also examined 4 and 12 carbon atoms on a 2$\times$2 cell, but only with the Brenner potential. Although Brenner potential gives wrong energy ranking, the true ground states were also found as low-energy configurations in the search. 
With this encouragement, we proceed to the case of more than one atomic types and variable stoichiometry.

\subsection{MgO (111)}
\begin{figure}[!htb]
\epsfig{file=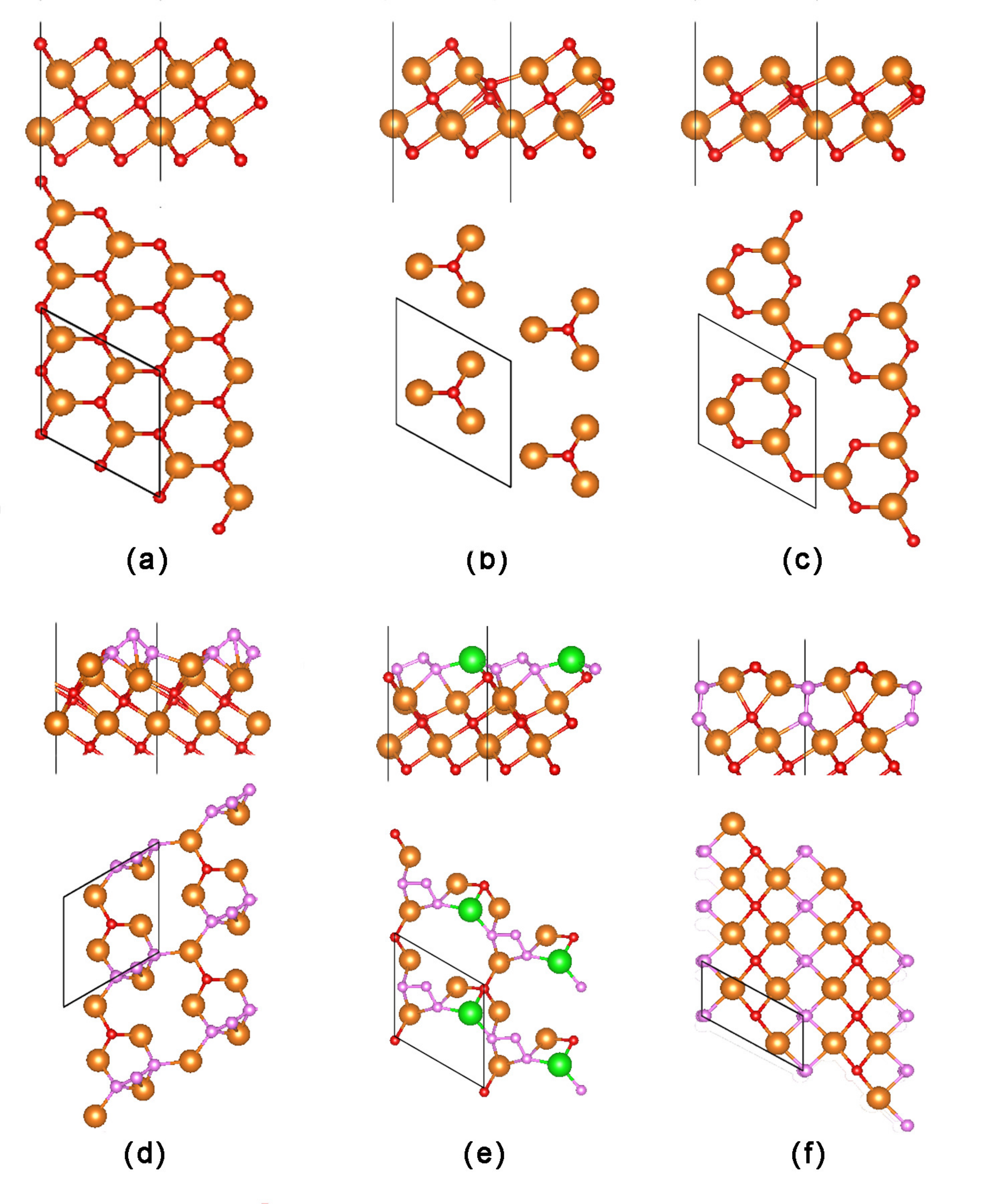, width=0.51\textwidth}
\caption{Side views (top, two layers only) and top views (bottom) of MgO (111) surfaces. (a) Unreconstructed surface; (b) 2$\times$2 O-terminated octupolar model; (c) 2$\times$2 $\alpha$-Mg model; (d) 2$\times$2 O$_3$ model; (e) 2$\times$2 Mg-O$_3$ model; and (f) 2$\times$1 O$_2$ model. The $\alpha$-Mg model can be derived by simply removing the topmost O atom from O-terminated octupolar model. Large spheres: magnesium; small spheres: oxygen. The topmost O$_3$ trimers in (d), O$_3$-Mg-O$_3$ chains in (e) and peroxo O$_2$ species in (f) are highlighted by different colors.}
\label{MgO}
\end{figure}

Magnesium oxide is one of the most abundant materials in planetary mantles, and also an archetypal ionic crystal with a rocksalt type (B1) structure. However, surfaces of B1-MgO present a challenging problem, especially for the polar (111) surface. An ideal bulk termination of MgO (111) surface contains alternating planes of Mg$^{2+}$ and O$^{2-}$ ions, allowing a surface dipole along the [111] direction, leading to extremely unstable surface configuration. To stabilize this polar (111) surface, there must be valence compensation to cancel the surface dipole, achieved either by rearrangement of surface species or adsorption of some foreign species. Ciston {\it et al.} \cite{Ciston-PRB-2009, Ciston-Surf-2010} investigated the coverage by hydroxyl (OH), combining transmission electron microscopy (TEM) and DFT calculations. They proposed several models of water-driven reconstructions. Regarding reconstruction of the clean surface, Wolf \cite{Wolf-PRL-1992} proposed a 2$\times$2 reconstruction of B1-NaCl ͑(111) surface based on valence-neutral octupolar units where 3/4 ions of the top layer and 1/4 ions of the second layer are missing, thus forming O$_1$-Mg$_3$-O$_4$ (O-terminated) or Mg$_1$-O$_3$-Mg$_4$ (Mg-terminated) surface stoichiometry. However, octopolar reconstructions are not fully consistent with the experimental data. Finocchi et al \cite{Finocchi-PRL-2004} found a Mg-terminated phase (designated as $\alpha$-Mg phase) with a 2$\times$2 repeating unit to be favored in O-poor conditions. They further proposed a combination of O-oct and $\alpha$-Mg which could fit the diffraction data. Using transmission high energy electron diffraction (THEED), Plass et al \cite{Plass-PRL-1998} reported another major class of reconstructions on MgO (111), which is based on cyclic oxygen trimer units (2$\times$2, $\sqrt{3}\times\sqrt{3}$, 2$\sqrt{3} \times 2\sqrt{3}$). DFT calculations \cite{Zhang-JPCC-2008} indicate that 2$\times$2 O-octopolar reconstruction is dominantly stable at a wide oxygen chemical potential range. $\alpha$-Mg is favored under Mg-rich conditions. However, the cyclic ozone-2$\times$2 is calculated to have much higher surface energy in the whole chemical potential range. The huge disagreement between simulation and experiment suggests that the structural model might need to be revised. 

\begin{figure}[!htb]
\epsfig{file=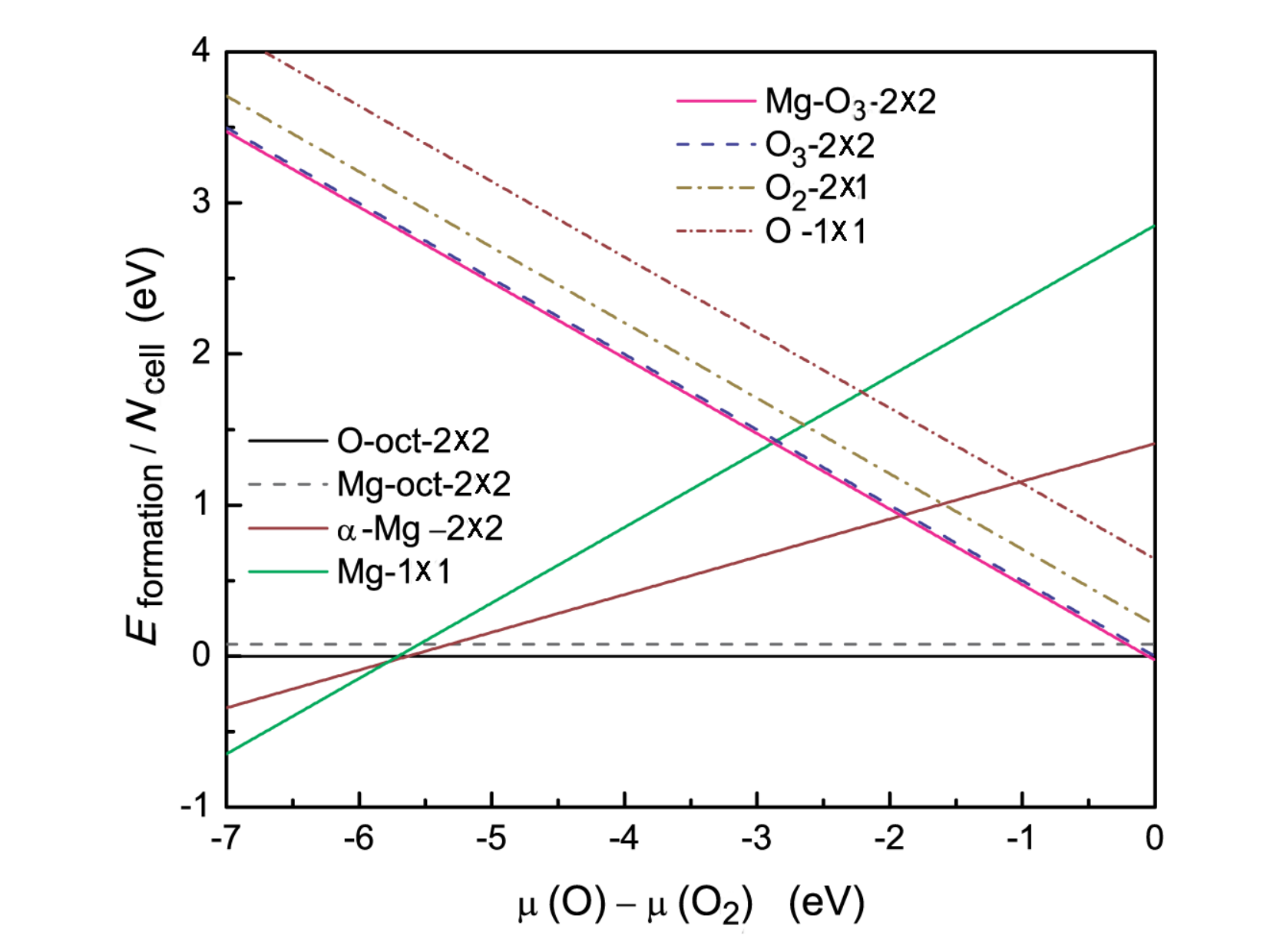, width=0.5\textwidth}
\caption{Zero temperature phase stability diagram for the proposed reconstructions on MgO (111) surface.}
\label{MgO-diagram}
\end{figure}

We used an EA search for the MgO (111) surface with fixed surface stoichiometry Mg$_1$O$_1$ per 1$\times$1 cell on the given substrate (O-terminated), and a variable cell size (from 1$\times$1 to 2$\times$2). Our search yields a low-energy reconstruction based on the O$_3$ trimer unit. As shown in Fig. \ref{MgO}d, the topmost surface is a periodic arrangement of trimers and single oxygen atoms, which is very similar to the model suggested by the Plass \cite{Plass-PRL-1998}. However, our search found a configuration with energy lower by -0.026 eV per 1$\times$1 cell. The side view of the this reconstruction (Fig. \ref{MgO}e) indicates that in each repeat unit, one Mg atom in the second layer jumps to the topmost layer, and thus leaves a Mg vacancy, displaying the same Mg$_3$ stoichiometry plane per 2$\times$2 cell as in the O-octupolar model and the $\alpha$-Mg model. The extra Mg atom in the top layer connects neighbouring O$_3$-trimers, forming periodic O$_3$-Mg-O$_3$ chains. Thus we call it Mg-O$_3$-2$\times$2. 

Another interesting reconstruction appearing in our search is a 2$\times$1 reconstruction containing the peroxo species [O$_2$]$^{2-}$. This motif has the same surface stoichiometry as Mg-O$_3$-2$\times$2; its surface energy is 0.23 eV per 1$\times$1 cell higher. Although not favored in (111) surface, it might exist in some other surfaces, such as the (100) surface \cite{Wang-Surf-1992}). This type of motif can provide a hint for experimental growth of the crystal MgO$_2$, which is calculated to be stable at high pressure \cite{Zhu-MgO-PCCP}. It indicates that novel chemistry under extreme conditions (such as high pressures) might be also achieved by careful exploiting of reduced dimensionality. 

Using an established way to correct for errors of the GGA functional, the calculated DFT energy of the O$_2$ molecule is considered to have a upward shift of about 1.05 eV \cite{Mellan-JCP-2012}. Including this empirical correction, we construct the phase diagram according to Eq. \ref{eq:E2}. The phase diagram is very similar to the diagram presented in Ref. \cite{Zhang-JPCC-2008}, except that the proposed models in our search have significantly lower energy than previously models with the same surface stoichiometry (Mg$_1$O$_1$). In particular, the Mg-O$_3$-2$\times$2 reconstruction is calculated to be stable at a narrow range of chemical potential when O is extremely rich ($\mu$(O)-$\mu$(O$_2$) $\ge$ 0.06 eV). Given that the low energy feature, it is very likely that these reconstructions could be observed under certain experimental protocols.

\subsection{GaN $(10\overline{1}1)$ with and without oxygen}
Applications so far use only fixed stoichiometry searches. To prove the possibility of automated exploring the whole stoichiometric space, we perform a study on the semipolar GaN $(10\overline{1}1)$ surfaces, which allows both variable surface atoms and variable cell size (restricted to a $2\times 2 $ or smaller surface cell). We also study the reconstructions on the semipolar GaN $(10\overline{1}1)$ surfaces in the presence of oxygen.

\begin{figure*}
\epsfig{file=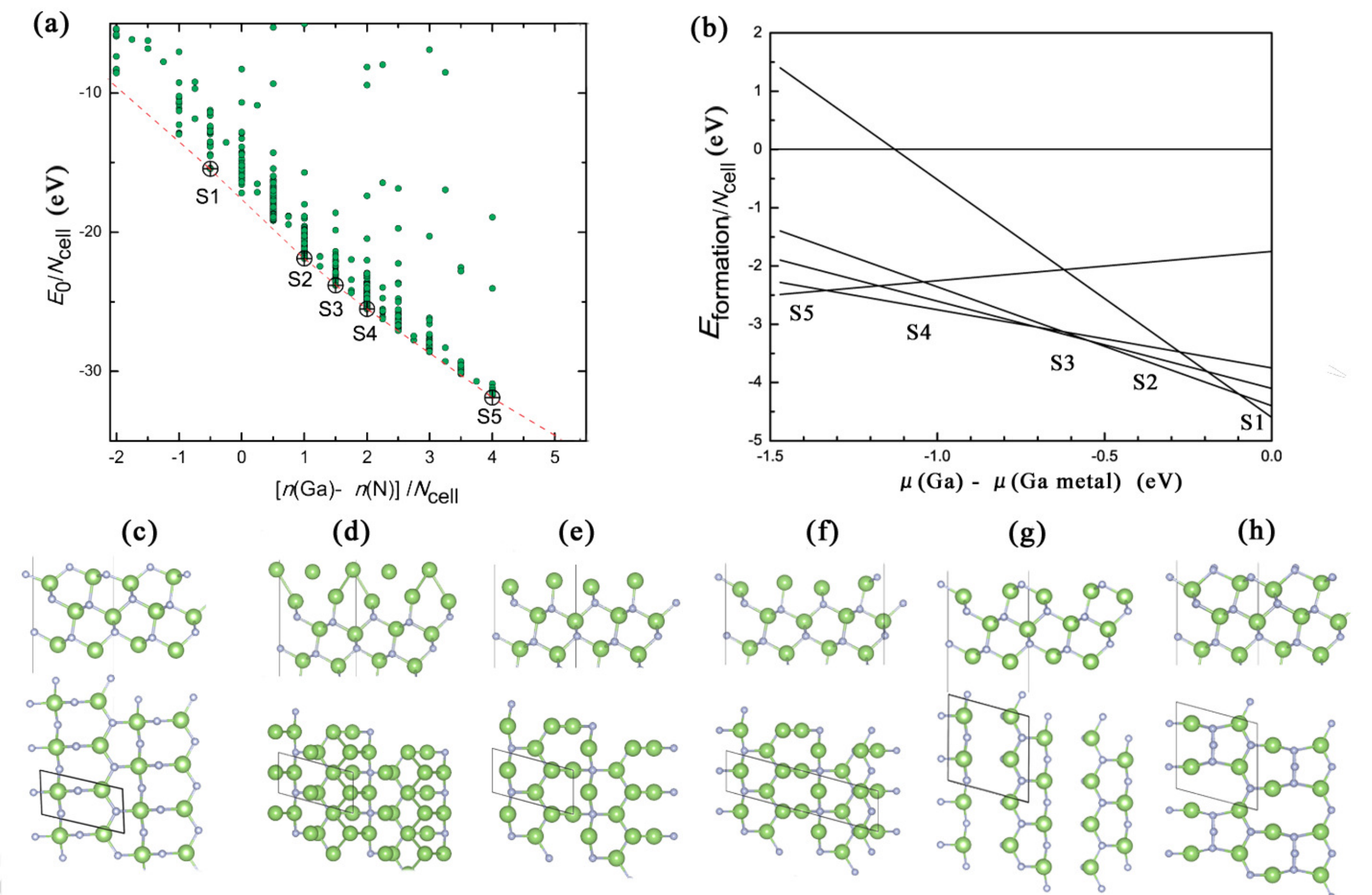, width=0.9\textwidth}
\caption{(a) Energies of GaN $(10\overline{1}1)$ surface reconstructions explored by the EA search; (b) the corresponding zero temperature phase stability diagram up to 2 $\times$ 2 supercell reconstructions. Side view and top view of (c) the cleaved GaN $(10\overline{1}1)$ surface and its lower-energy reconstructions found by the EA search; (d) S1 configuration: (1$\times$1) Ga-bilayer; (e) S2 configuration: (1$\times$1) Ga-monolayer; (f) S3 configuration: (2$\times$1)-1.5 N vacancy; (g) S4 configuration: (2$\times$1)-1 N vacancy; (h) S5 configuration: (2$\times$1) N-trimer. Large spheres, gallium; small spheres, nitrogen.}
\label{convex_hull}
\end{figure*}
Fig. \ref{convex_hull}a gives the convex hull diagram from our EA search, and the five corresponding stable structures. Compared to the cleaved surface (Fig.~\ref{convex_hull}c), structure S1 has two Ga adlayers.  Structure S2 has one Ga adlayer.  Structure S3 has the top N and half of the second N removed.  Structure S4 has only the top N removed.  Structure S5 has an additional N at the bridging position of the two top N atoms.  The first 4 structures have been found by Akiyama \textit{et al} \cite{Akiyama-PRB-2010}.  Structure S5 with N$_3$ trimers, was not reported before and is not intuitively obvious. This demonstrates the power of the automated searching by the EA.  An analogous Se-trimer feature has been predicted to be stable on ZnSe (100) reconstructions at Se-rich conditions \cite{Garcia-APL-1994}. Compared to the cleaved surface, the stable structures can be constructed through either vacancies or adatoms.  No more complicated reconstructions have been found.  For the Ga-rich conditions, there are many intermediate structures between structure S1 and S2, lying very close to the convex hull (see Fig.~\ref{convex_hull}).  This can be explained by the weak directionality of the Ga-Ga bonds. These structures may appear under finite temperatures.  For the N-rich conditions, however, most structures are much higher above the convex hull, due to the strong covalent bonding nature of N-N interactions.

The features possessed by the stable configurations are: i) Ga layers;  ii) N vacancies;  iii) N trimers. Another common feature is N$_2$-dimers \cite{Garcia-APL-1994}. These do not lie on the convex hull, but frequently appear as energetically competitive in our EA search.

The electron counting (EC) rule \cite{Pashely-PRB-1989, Zhang-PRL-2004, Zhang-PRL-2006} emerges from the study of semiconductor surfaces. This rule, although not exact, is very useful in comparing surface energies, estimating surface charges and guessing surface reconstructions. It states that the surface structures with filled dangling bonds on the electronegative elements and empty dangling bonds on the electron positive elements are the lowest in energy. The surface then is likely to be semiconducting. Surfaces not satisfying this rule will be metallic and higher in energy. None of the stable reconstructions found here satisfy the electron counting rule. Large reconstructions may satisfy the EC rule. Indeed, Lee and Kim \cite{Lee-PRB-2011} demonstrated that low-energy semiconducting surfaces can be achieved by large reconstructions, complying with the EC rule by constructing models based on N dimer and dimer-vacancy. Due to the limitation of computational power, we are only able to predict the reconstructions up to 2 $\times$ 2 surface cells. But our results provide a good starting point for predicting large reconstructions.

\begin{figure}
\epsfig{file=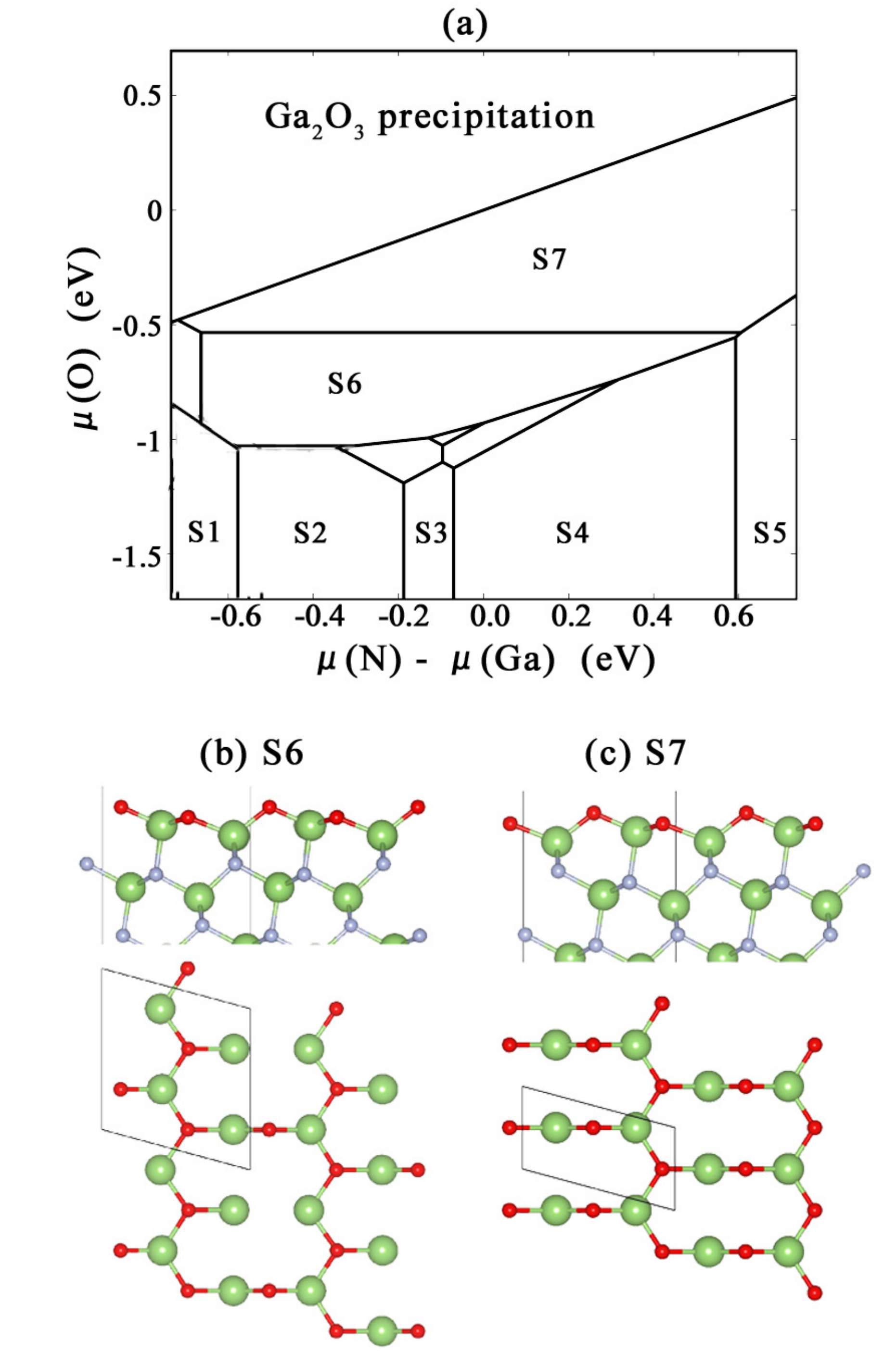, width=0.40\textwidth}
\caption{(a) Phase diagram of the GaN $(10\overline{1}1)$ surface with oxygen present up to 2 $\times$ 2 supercell reconstructions. Important reconstructions are labeled from S1 to S7. Structures S1 to S5 are shown in Fig. \ref{convex_hull}. Side view and top view of structures S6 and S7 are shown in (b) and (c). Large balls, gallium; small gray balls, nitrogen; small red balls, oxygen.}
\label{phase}
\end{figure}

Let us now employ the same method to search for reconstructions of GaN $(10\overline{1}1)$ surfaces with added oxygen. The formation energy can be defined as
\begin{equation}
E_\text{formation} = E_0 - \mu(\text{Ga}) [n(\text{Ga}) - n(\text{N})] - \mu(\text{O}) n(\text{O}).
\label{eq:E3}
\end{equation}
The oxygen chemical potential $\mu$(O) is a new variable. It has no lower bound, but would possibly reach the upper bound by the constraints based on the possible O$_2$, N$_x$O$_y$ molecular gas and  Ga$_2$O$_3$ solid:
\begin{equation}
\begin{split}
2\mu(\text{O}) &\leq ~G(\text{O}_2),~~~ \\
2\mu(\text{Ga}) + 3\mu(\text{O}) &\leq ~G(\text{Ga}_{2}\text{O}_{3}),\\
x\mu(\text{Ga}) + y\mu(\text{O}) &\leq ~G(\text{N}_{x}\text{O}_{y}).
\end{split}
\end{equation}
Therefore, the convex hull has two dimensions in $n(\text{O})$ and [$n(\text{Ga}) - n(\text{N})$]. Fig. \ref{phase}a shows the two-dimensional phase diagram in space of $\mu(\text{O})$ and [$\mu(\text{N}) - \mu(\text{Ga})$], transformed from the convex hull found in our EA search. The two new major reconstructions are structure S6 and S7. Compared to the cleaved surface, structure S6 has half of the top N removed, half of the top N and all of the second N replaced by O. Structure S7 has the top two N replaced by O. Similar reconstructions for the $(10\overline{1}1)$ surface have been reported \cite{Northrup-PRB-2006}. These reconstructions are favored because of the strong Ga-O bond and the electron counting rule. For example, the simple cleaved surface has 3 nitrogen dangling bonds per surface cell. Because each N dangling bond needs 3/4 electron to become filled, the surface needs 9/4 electrons per surface unit cell to satisfy the electron counting rule. For structure S7, substituting 2 N with 2 O introduces 2 extra electrons, making the electron counting rule almost satisfied.

\section{Conclusions}
In summary, we present a specially designed method to automatically explore low energy surface reconstructions with variable surface atoms and reconstruction cells. The power of this approach is illustrated by application to diamond (100)/(111) surfaces. Applying to MgO (111) and GaN-O $(10\overline{1}1)$ surfaces, we find several new low-energy reconstruction models. All of the systems investigated in this paper show quite complex surface reconstruction behaviors, indicating that an automated search around the potential energy landscape can be very helpful for us to understand surface chemistry. We demonstrate our method of predicting surface reconstruction based on evolutionary approach can be generally applied to any binary and even multi-component systems.

\begin{acknowledgments}
Calculations were performed at the supercomputer of Center for Functional Nanomaterials, Brookhaven National Laboratory, which is supported by the U.S. Department of Energy, Office of Basic Energy Sciences, under contract No. DE-AC02-98CH10086, and at joint Supercomputer Center of the Russian Academy of Sciences and on Skif supercomputer (Moscow State University). This work is funded by DARPA (grants W31P4Q1210008 and W31P4Q1310005), National Science Foundation (No. EAR-1114313) and DOE (No. DE-FG02-08ER46550). QZ thanks W-B Zhang for providing data on the MgO (111) surface. PBA and LL thank the SWaSSiT collaboration, and in particular, M. V. Fernandez-Serra, M. S. Hybertsen, and J. T. Muckerman, for help.
\end{acknowledgments}

\bibliography{reference}

\end{document}